\documentclass[12pt]{article}
\usepackage{amssymb,amsmath,graphicx}
\makeatletter

\@addtoreset{equation}{section}
\def\section{\@startsection {section}{1}{\z@}{-2.5ex plus -1ex minus
 -.2ex}{1.3ex plus .2ex}{\large\bf}}
\def\subsection{\@startsection{subsection}{2}{\z@}{-2.25ex plus%
 -1ex minus -.2ex}{0.5ex plus .2ex}{\bf}}

\advance \voffset by -0.8in
\advance \hoffset by -0.6in
\def\cd{\!\cdot\!}

\def\MM{F}
\def\Pp{Q'}
\def\P{Q}
\def\cP{P}
\def\bPp{{\mbox{\boldmath $\Pp$}}}

\def\bp{{\mbox{\boldmath $p$}}}
\def\bP{{\mbox{\boldmath $P$}}}
\def\bq{{\mbox{\boldmath $q$}}}

\def\bJ{{\mbox{\boldmath $J$}}}
\def\bP{{\mbox{\boldmath $\P$}}}

\def\bm{{\mbox{\boldmath $m$}}}
\def\bv{{\mbox{\boldmath $v$}}}
\def\bw{{\mbox{\boldmath $w$}}}

\def\bpm{\begin{pmatrix}}
\def\epm{\end{pmatrix}}
\newcommand{\cm}{\mathfrak{m}}  
\newcommand{\cg}{\mathfrak{g}}
\newcommand{\ch}{\mathfrak{h}}

\newcommand{\RR}{\mathbb{R}}
\newcommand{\CC}{\mathbb{C}}
\newcommand{\tr}{{\rm tr}}
\newcommand{\ad}{{\rm ad}}
\newcommand{\Ad}{{\rm adj}}
\newcommand{\id}{{\rm id}}
\def\bee{\begin{equation}}
\def\eee{\end{equation}}
\def\bea{\begin{eqnarray}}
\def\eea{\end{eqnarray}}
\newcommand{\bra}[1]{\langle\,#1,\cdot\rangle}

\newcommand{\ket}[1]{#1}

\newtheorem{theorem}{Theorem}[section]

\newtheorem{lemma}[theorem]{Lemma}

\def\lrbicross{{\blacktriangleright\!\!\!\triangleleft}}

\begin{document}
\parskip 4pt
\parindent 8pt
\begin{flushright}
EMPG-13-12
\end{flushright}

\begin{center}

{\Large \bf Classical $r$-matrices  via semidualisation }

\baselineskip 20 pt

\vspace{.2cm}

{ \bf Prince~K.~Osei } \\
Department of Mathematics  \\
 University of Ghana, 
PO Box LG 25,
Legon, Ghana \\
{pkosei@ug.edu.gh}

\vspace{.2cm}

 {\bf  Bernd~J.~Schroers}   \\
Department of Mathematics and Maxwell Institute for Mathematical Sciences \\
 Heriot-Watt University, 
Edinburgh EH14 4AS, United Kingdom \\ 
{b.j.schroers@hw.ac.uk}

\vspace{0.3cm}

{24 August 2013, udpated 8  October  2013}
\baselineskip 16 pt

\end{center}

\begin{abstract}
\noindent We study the interplay between  double cross sum decompositions of a given Lie algebra and classical $r$-matrices for its semidual. For a class of Lie algebras which can be obtained  by a process of generalised complexification we derive an expression for  classical $r$-matrices
of the semidual Lie bialgebra in terms of the data which determines the  decomposition of the original Lie algebra.
Applied to the local isometry Lie algebras arising in three-dimensional gravity, decomposition and semidualisation yields the main class of   non-trivial  $r$-matrices for the  Euclidean and  Poincar\'e group in three dimensions. In addition, the construction  links the $r$-matrices with the  Bianchi classification of  three dimensional  real Lie algebras. 
\end{abstract}

%\centerline{PACS numbers: 04.20.Cv, 02.20.Qs, 02.40.-k}

\section{Introduction}

In classical geometry and physics, the geometry of a  homogeneous space  and its  isometry group  mutually determine each other.  
Any deformation of either the geometry into a non-commutative version or of the isometry group into a Hopf algebra should  preserve as much of this interplay as possible. One way of achieving this, at least at the algebraic level, is to combine the space   and its  isometry into one algebraic object.  In physics terms, such an algebra should combine position coordinates for space  with  generators of rotations  and  translations, i.e.,  with angular momenta and momenta. 

Momenta and positions are in duality and rotations act on both. This is manifest in three important subalgebras: the Heisenberg algebra generated by positions and momenta, the  isometry algebra  generated by angular momenta and momenta and finally the algebra of angular momenta  and positions, which one may interpret as an isometry algebra of momentum space.  In these general terms, semiduality is  a bijection between the  isometry algebras  of space and momentum space.
%, i.e. the replacement of  the  momentum algebra  with the dual position algebra in the context of the rotation-momentum-position algebra. 

A mathematically precise version of these ideas was  formulated by Majid   in terms of  Hopf algebras \cite{Majidbicross} or, infinitesimally, Lie bialgebras \cite{Majid90}, and is summarised in the textbook \cite{Majid}. Majid's approach is in turn inspired by  Born's proposal of a reciprocity between momenta and positions \cite{Born}.  Here we will use Majid's framework,  working at the level of  Lie bialgebras. Thus, non-commutativity of a space is parametrised by Lie brackets of  position coordinates and its curvature by their co-commutator. In the dual Lie bialgebra, commutators and co-commutators are swapped so that curvature of space is also captured in the commutators of momenta;  similarly curvature of  momentum space is captured in the co-commutators of momenta or the commutators of positions.   

Curved, homogenous geometries and  the associated Lie brackets of isometry algebras are a standard topic in classical geometry and physics. Curved, homogeneous momentum spaces and the associated co-commutators of momenta are less familiar but play  a central role in 3d (quantum) gravity (see \cite{SchroersCracow} for a review),  in so-called $\kappa$-deformations of the Poincar\'e Lie algebra \cite{LNRT,LNR,MR} and more recently in the discussion about relative locality \cite{relloc}.  In  this paper we exploit  our familiarity with curvature and non-commutativity on one hand to enhance our understanding of co-commutativity on the other. We consider a family of Lie algebras, thinking of them as  rotation-position algebras or infinitesimal isometries of  momentum space,   and systematically compute their semidual bialgebras. The semiduals have non-trivial co-commutators
 which are co-boundary, i.e.,  given  by a  classical $r$-matrix. We derive  a  formula for the $r$-matrix in terms of the map which characterises the split of the original Lie algebra into rotation generators and positions.

 Specialising to three dimensions,   the semiduals of our family of Lie bialgebras have the Lie brackets of the  3d  Euclidean or  Poincar\'e  groups.
Lie bialgebra structures on these Lie algebras are necessarily co-boundary and the possible $r$-matrices were classified a while ago by Stachura in \cite{Stachura}.  Denoting angular momenta by $J_a$ and momenta by $P_a$, the list found by Stachura includes $r$-matrices which only contain rotation or Lorentz generators (these are just the standard $r$-matrices for  $sl(2,\RR)$), trivial solutions which only contain the commuting momentum generators, and then a more interesting and complicated list of solutions of the mixed form $r=R^{b}_{\;\;a} \cP^a \wedge J_b$.  Our approach via semiduality reproduces all these mixed solutions and relates them to the Lie brackets of the original algebra. In particular, we are able to relate these $r$-matrices to the Bianchi classification of three-dimensional Lie algebras.

The work reported here is closely related to our  previous paper \cite{OS1}  in which we studied  the semidual Hopf algebras of the universal enveloping algebras of the isometry Lie algebras arising in 3d gravity. Here we generalise and extend the results of \cite{OS1}  at the infinitesimal,  Lie bialgebra level. However, in comparing the current paper with \cite{OS1} and also with previous work on semiduality in 
 3d quantum gravity \cite{FNR,MajidSchroers}  the reader should be aware of a  possible confusion between different interpretations of semiduality. Here we interpret semiduality essentially as the exchange of position and momentum generators, as outlined above.  In \cite{MajidSchroers,OS1}, by  contrast,  pairs of semidual Hopf algebras are both thought of as `rotation-momentum' algebras, but semiduality exchanges the regimes of 3d quantum gravity where they play the role of quantum isometry groups.

The paper is organised as follows. In Sect.~2  we introduce the family of Lie algebras whose direct sum decomposition we would like to study. They are necessarily even-dimensional and obtained from a given real Lie algebra $\cg$ by a process of generalised complexification.  We  parametrise decompositions of  Lie algebras in the original family  as double cross sums with $\cg$ as one of the factors in terms of a map $F:\cg \rightarrow \cg$ and derive a quadratic Lie-algebraic relation  for $F$ which ensures that it does indeed define a double cross sum.  The main result of this section is that such maps  also characterise classical $r$-matrices of the semidual Lie bi-algebra, and that the quadratic relation obtained as factorisation condition is equivalent to the modified classical Yang-Baxter equation for the semidual Lie bi-algebra.  In Sect.~3 we specialise to the case where $\cg$ is the Lie algebra of  either the rotation group in Euclidean 3-space or the Lorentz group in three spacetime dimensions. The resulting family of complexified Lie algebras are the local isometry groups of 3d gravity \cite{SchroersCracow}, but here we think of them as the Lie algebras of rotations or Lorentz boosts together with position coordinates, so that, in keeping with the general philosophy explained above,  their semiduals, which have the Lie algebra structure of the  3d Euclidean or Poincar\'e Lie algebra,  can be interpreted as spacetime symmetries. Using the isomorphism between $\cg\wedge\cg$ and $\cg$ in this case, we reformulate the quadratic Lie-algebraic relation for $F$ as a quadratic equation for the linear map $F$,  and, using rotational or Lorentz symmetry,  project out three equations which have to be satisfied independently. Solving these equations in Sect.~4 we recover a family of $r$-matrices for the Euclidean  and Poincar\'e Lie algebras first found by Stachura \cite{Stachura}, and show that they are in one-to-one correspondence with the types in   the Bianchi classification of three-dimensional Lie algebras bar the Heisenberg algebra. Sect.~5 contains a summary of our results in tabular form and a discussion. 

\section{A general theorem on double cross sums and $r$-matrices for their semidual}

\subsection{Direct sum decompositions of complexified Lie algebras }
Consider  an {\em } arbitrary $n$-dimensional real Lie algebra $\cg$  with generators $\{J_a\}_{a=1,\ldots,n}$ and brackets 
\bee
[J_a,J_b]=f_{ab}^{\;\;\;c}J_c.
\eee
Here and in the following we use the  Einstein summation convention. 
Picking a real number $\lambda$, one can associate to this Lie algebra the following $2n$-dimensional real Lie algebra $\cg_\lambda$
with generators $\{J_a\}$, additional generators $\P_a$, $a=1,\ldots,n$,  and brackets of the Cartan form
\bee
\label{genisolie}
[J_a,J_b]=f_{ab}^{\;\;\;c}J_c, \quad [\P_a,J_b]=f_{ab}^{\;\;\;c}\P_c, \quad [\P_a,\P_b]=\lambda f_{ab}^{\;\;\;c}J_c.
\eee
For $\lambda =-1$ this is the usual complexification $\cg\otimes \CC$, for $\lambda =0$ it is the  semidirect product $\cg \ltimes \RR^n$ and for $\lambda =1$ it is isomorphic to $\cg \oplus \cg$.  One can unify these three cases by considering the Lie algebra $\cg_\lambda$ as a generalised complexification  with a formal parameter $\theta$ which satisfies $\theta^2=-\lambda$
and the identification $\P_a=\theta J_a$, see \cite{Meusburger} and \cite{MS}. We will not emphasise this viewpoint in the following, but will make use of the linear  map $\theta$ defined via
\bee
\theta: \cg_\lambda \mapsto \cg_\lambda, \quad J_a\mapsto \P_a, \quad  \P_a\mapsto \lambda J_a\quad \text{for} \;a=0,\ldots,n.
\eee

Lie algebras like \eqref{genisolie} arise in physics, particularly in three spacetime dimensions. In that context, the generators $J_a$ are rotation- or Lorentz generators,  the  
$\P_a$ are usually interpreted as components of momentum, i.e., as generators of spacetime translations, and the constant $\lambda $  is related to the cosmological constant.  In the current context we also think of the $J_a$ as generalisations of rotation or Lorentz generators. However, when we apply semiduality we switch from the generators $\P_a$ to generators of the dual vector space, which we then  interpret as momentum space.  From that  point of view, the generators $\P_a$ should be thought of as position coordinates, and    $\lambda$ as a  real constant related to the curvature of momentum space \cite{MajidSchroers,OS1}

We are interested in all decompositions of $\cg_\lambda$  as double cross sum  $\cg_\lambda =\cg \bowtie \cm$
of  $\cg$ and a second $n$-dimensional Lie algebra $\cm\subset \cg_\lambda$.  We can assume without loss of generality that the generators of $\cm$ are of the form 
\bee
\label{genguessgen}
\Pp_a=\P_a +\MM^b_{\;\;a}J_b, \quad a=1,\ldots,n,
\eee
for a real $n\times n$ matrix $\MM^b_{\;\;a}$.  As explained in \cite{Majid},  the most general form of the brackets in such a double cross sum is 
\bee
\label{gennisolie}
[J_a,J_b]=f_{ab}^{\;\;\;c}J_c, \quad [\Pp_a,J_b]=f_{ab}^{\;\;\;c}\Pp_c +L_{ab}^{\;\;\;c} J_c, \quad [\Pp_a,\Pp_b]=g_{ab}^{\;\;\;c}\Pp_c.
\eee
If the matrix elements $\MM^b_{\;\;a}$ can be found so that the  generators $\Pp_a$ close under Lie brackets then
 we can express the structure constants $L_{ab}^{\;\;\;c}$ and $g_{ab}^{\;\;\;c}$ in terms of $\MM$ and $f_{ab}^{\;\;\;c}$ as follows:
\begin{align}
\label{step1}
g_{ab}^{\;\;\;c} = f_{ad}^{\;\;\;c}\MM^d_{\;\;b}+ \MM^d_{\;\;a}f_{db}^{\;\;\;c}, \qquad 
 L_{ab}^{\;\;\;c}= \MM^d_{\;\;a}f_{db}^{\;\;\;c}-\MM^c_{\;\;d}f_{ab}^{\;\;\;d}.
\end{align}
To determine the condition on the generators \eqref{genguessgen} to form a Lie subalgebra
we think of $\MM^b_{\;\;a}$ as the matrix relative to the basis $\{J_a\}_{a=1,\ldots,n}$  of a linear map 
\bee
\MM:\cg \rightarrow \cg, \qquad  X=X^a J_a\mapsto \MM^{b}_{\;\;a}X^a J_b,
\eee
Then we define the map
\bee
\label{Idef}
I:\cg \rightarrow \cg_\lambda, \qquad X\mapsto \theta X + \MM(X), 
\eee
so that  the sought-after generators in \eqref{genguessgen} are 
\bee 
\label{genguessgenn}
\Pp_a=I(J_a), \qquad a=1,\ldots,n.
\eee
% we can write the bracket in $\cm$ as 
%\bee
%[I(X),I(Y)]= I([M(X),Y]+ [X,M(Y)]),
%\eee
%and the mixed brackets as 
%\bee
%[I(X),Y]= I([X,Y])+
%([M(X),Y]-M [X,Y]).
%\eee
Using this notation, we show the following. 
\begin{lemma}
The  condition on $\MM:\cg\rightarrow \cg$ for the generators \eqref{genguessgenn} to form a Lie subalgebra of $\cg_\lambda$ is 
\bee
\label{Mcond} 
 [\MM(X),\MM(Y)]-\MM([X,\MM(Y)]+[\MM(X),Y]) =-\lambda[X,Y] \quad \forall \; X,Y\in\cg.
\eee
\end{lemma}
{\bf Proof} \;  In terms of the map \eqref{Idef} we need to find the condition under which the commutator of the two elements
of the form $I(X)$ and $I(Y)$  lies in the image of $I$. One computes
\bee
[I(X),I(Y)]= \lambda[X,Y] +[\MM(X),\MM(Y)] +\theta([X,\MM(Y)]+ [\MM(X),Y]).
\eee
For this to be of the form $I(Z)=\theta Z + \MM(Z)$ we require
\bee
\MM([X,\MM(Y)]+ [\MM(X),Y])=\lambda[X,Y] +[\MM(X),\MM(Y)],
\eee
as claimed. \hfill $\Box$

Note that, when the map $\MM$ solving  \eqref{Mcond} can be found,  then the map $I$ is a bijection $\cg\rightarrow\cm$

\subsection{Semidual  Lie  bialgebras and  their  classical r-matrices}

The  process of semidualisation is defined for  Lie bialgebras which are double cross sums, see \cite{Majid} for a systematic treatment. 
We are going to apply it to the Lie bialgebras $\cg\bowtie\cm$, with the Lie algebra structure already discussed and trivial co-commutator. In the semidual Lie bialgebra, the Lie algebra $\cm$ is replaced  by its dual Lie bialgebra $\cm^*$. In our case,  $\cm$ has trivial co-commutators so that the Lie algebra structure of $\cm^*$ is abelian. Following the notation of \cite{Majid}, we denote the semidual  Lie bialgebra of $\cg\bowtie\cm$ by $\cm^*\lrbicross\cg.$ In order to give its commutators and co-commutators, we introduce generators $\cP^a$, $a,=1,\ldots, n$ of $\cm^*$ which are dual to the generators $\Pp_a$ of  $\cm$, i.e.,  they satisfy 
 \bee \cP^a(\Pp_b)=\delta^a_b.
 \eee 
Then, in terms of  the  generators $\{J_a,\cP^b\}_{a,b=1,\ldots, n} $ of the semidual Lie bialgebra,  the commutators are 
\bee
\label{semid}
[J_a,J_b]=f_{ab}^{\;\;\;c}J_c, \quad [J_a,\cP^b]=-f_{ac}^{\;\;\;b}\cP^c, \quad [\cP^a,\cP^b]=0,
\eee
or
\[
[J_a,J_b]=f_{ab}^{\;\;\;c}J_c, \quad [\cP^a,J_b]=f_{bc}^{\;\;\;a}\cP^c, \quad [\cP^a,\cP^b]=0.
\]
Applying  the general formulae in Sect.~8.3 of  \cite{Majid}, the    co-commutators  in our case come out as 
\begin{align}  
\label{dualco}
\delta (\cP^a)  = g_{cb}^{\;\;\;a} \cP^c \otimes \cP^b , \qquad 
\delta (J_a)  = L_{ba}^{\;\;\;c} \left( J_c\otimes \cP^b -\cP^b \otimes J_c \right).
\end{align}

\begin{theorem} 
For each double cross sum decomposition of the Lie algebra $\cg_\lambda$ in the form $\cg\bowtie\cm$ and with generators of $\cm$ given in terms of the map  $\MM:\cg \rightarrow \cg$ by \eqref{genguessgenn},   the semidual Lie bialgebra $\cm^*\lrbicross\cg$ has a co-commutator which is co-boundary, with classical $r$-matrix
\bee
\label{rmatrix}
r=P^a \wedge F( J_a) = \MM^{b}_{\;\;a}\cP^a \wedge J_b.
\eee 
 \end{theorem}
{\bf Proof}  \; Consider the following ansatz for an $r$-matrix of $\cm^*\lrbicross\cg$
\bee
\label{genr}
r=R^{b}_{\;\;a} \cP^a \wedge J_b.
\eee
For the associated co-commutators one finds 
\begin{align}
\delta(\cP^a)&= \left(-R^d_{\;\;b}f_{dc}^{\;\;\;a}+R^d_{\;\;c}f_{db}^{\;\;\;a}\right)\cP^c\otimes\cP^b \nonumber \\
\delta(J_a)&=\left(-R^d_{\;\;b}f_{ad}^{\;\;\;c} + R^c_{\;\;d}f_{ab}^{\;\;\;d}\right)\left(J_c\otimes \cP^b -\cP^b\otimes J_c\right)
\end{align}
so that we  obtain the relation
\begin{align}
 g_{ab}^{\;\;\;c} =R^d_{\;\;b}f_{ad}^{\;\;\;c}+R^d_{\;\;a}f_{db}^{\;\;\;c}, 
 \qquad
 L_{ab}^{\;\;\;c} &=R^d_{\;\;a}f_{db}^{\;\;\;c} -R^c_{\;\;d}f_{ab}^{\;\;\;d}.
\end{align}
Comparing with  the expressions  for  $g_{ab}^{\;\;\;c}$ and $ L_{ab}^{\;\;\;c}$ given in \eqref{step1}  we conclude that we can  reproduce them by choosing
\bee
\label{result1}
  R^a_{\;\;b}=\MM^a_{\;\; b}.
\eee
Finally, we determine a condition for \eqref{genr} to satisfy the 
 modified classical Yang-Baxter equation 
\bee
\label{mybe}
[[r,r]]+ \lambda \Omega =0,
\eee
where, for now,  $\lambda$ is an arbitrary real  constant and   $\Omega$ is the  invariant element 
\bee
\Omega= f_{ab}^{\;\;\;c}( \cP^a \otimes \cP^b  \otimes J_c - \cP^a \otimes J_c \otimes \cP^b + J_c \otimes \cP^a  \otimes \cP^b).
\eee
Inserting \eqref{genr} into 
\bee
[[r,r]] = [r_{12},r_{13}]+ [r_{12},r_{23}] + [r_{13},r_{23}]
\eee
yields three terms,
\bee
 \left(R^b_{\;\;a} R^c_{\;\;d}f_{be}^{\;\;\;d} - R^b_{\;\;e} R^c_{\;\;d}f_{ba}^{\;\;\;d}+R^b_{\;\;e} R^d_{\;\;a}f_{bd}^{\;\;\;c}\right) \cP^e \otimes \cP^a  \otimes J_c
\eee
and similar terms proportional 
 to $\cP^a \otimes J_c \otimes \cP^b$, and $J_c \otimes \cP^a  \otimes \cP^b$.
 We  deduce that the modified Yang-Baxter equation  \eqref{mybe} is equivalent to
\bee
R^b_{\;\;a} R^c_{\;\;d}f_{be}^{\;\;\;d} - R^b_{\;\;e} R^c_{\;\;d}f_{ba}^{\;\;\;d}+R^b_{\;\;e} R^d_{\;\;a}f_{bd}^{\;\;\;c}+\lambda f_{ea}^{\;\;\;c}=0.
\eee
However, this is simply the matrix form of the equation \eqref{Mcond} for the map $\MM$. Thus,  substituting  \eqref{result1} into \eqref{genr} we  indeed   obtain a solution of the modified classical Yang-Baxter equation.  
\hfill $\Box$

\section{Application to the isometry Lie algebras of 3d gravity}

\subsection{Special properties in three dimensions}

We now turn to solutions of the factorisation condition for isometry Lie algebras arising in 3d gravity. In the following $\cg$ stands for either $so(3)$ or $so(2,1)$, with generators $J_a$,  $a=0,1,2$. This range of indices is unconventional in the Euclidean setting but  well-adapted to the more intricate  Lorentzian situation. We write   $\eta_{ab}=\eta^{ab}$   for either the Euclidean metric diag$(1,1,1)$ or the Lorentzian metric diag$(1,-1,-1)$, and use it lower or raise indices. The Lie brackets of $\cg$ are  then
\bee
\label{so123}
[J_a,J_b]=\epsilon_{abc}J^c,
\eee
where we  adopt the convention $\epsilon_{012}=\epsilon^{012}=1$. 
The invariant inner product 
\bee
\label{innerprod}
\langle J_a,  J_b\rangle =\eta_{ab}
\eee
on $\cg$ plays an important role in our analysis. 
In the Lorentzian case, it is sometimes convenient to work with  normalised  raising and lowering operators
\bee
\label{newgens}
 N=\frac{1}{\sqrt{2}}(J_0+J_2) , \qquad \tilde N =\frac{1}{\sqrt{2}}(J_0-J_2),
\eee
whose names are chosen to reflect the fact that they are null (or lightlike) with respect to  \eqref{innerprod}:
\bee
\langle N,  N \rangle = \langle\tilde  N, \tilde N \rangle=0, \quad \langle N, \tilde N \rangle  =1.
\eee
Their commutators are 
\bee
 [\tilde N,N] = J_1, \quad [J_1,N]=N, \quad [J_1,\tilde N] = -\tilde N.
\eee

The Lie  algebra $\cg_\lambda$ is that of the  Poincar\'e, de Sitter or anti-de Sitter  group or their Euclidean analogues in three dimensions, with brackets
\bee
\label{isolie}
[J_a,J_b]=\epsilon_{abc}J^c, \quad [\P_a,J_b]=\epsilon_{abc}\P^c, \quad [\P_a,\P_b]=\lambda \epsilon_{abc} J^c.
\eee
As in the general case, we are looking for basis change 
\bee
\label{guessgen}
\Pp_a=\P_a +\MM^{b}_{\;\;a}J_b,
\eee
so that $\{J_0,J_1,J_2,\Pp_0,\Pp_1,\Pp_2 \}$ is  a basis  and   $\{\Pp_0,\Pp_1,\Pp_2 \}$ closes under Lie brackets, i.e., 
\bee
\label{EuP}
[J_a,J_b]=\epsilon_{ab}^{\;\;\;c}J_c, \quad [\Pp_a,J_b]=\epsilon_{ab}^{\;\;\;c}\Pp_c +L_{ab}^{\;\;\;c} J_c, \quad [\Pp_a,\Pp_b]=g_{ab}^{\;\;\;c}\Pp_c.
\eee
In this case, the semidual Lie algebra \eqref{semid}  has the brackets
\bee
\label{eucllorentz}
[J_a,J_b]=\epsilon_{ab}^{\;\;\;c}J_c, \quad [J_a,\cP^b]=\epsilon_{ab}^{\;\;\;c}\cP^c, \quad [\cP^a,\cP^b]=0,
\eee
which are the brackets of the Euclidean Lie algebra in the case of Euclidean signature and those of the Poincar\'e Lie algebra in the case of Lorentzian signature. Every decomposition of \eqref{isolie} according to \eqref{guessgen} will therefore lead to a co-boundary Lie bialgebra structure on those Lie algebras.

In order to determine all possible solutions of the  condition \eqref{Mcond}, we will be making use of the invariant inner product  \eqref{innerprod} on $\cg$.
We write $\MM^t: \cg \rightarrow\cg $ for the transpose of a map $\MM:\cg \rightarrow \cg$ relative to  $\langle\;\;,\;\;\rangle$,  i.e.,  for the map which satisfies
\bee
\langle \MM^t (X),Y\rangle = \langle X, \MM (Y)\rangle \quad \forall X,Y\in \cg. 
\eee
We also need the fundamental identity
\bee
\label{epsid}
\epsilon_{abc}\epsilon^{efg} = \delta_a^e (\delta_b^f \delta_c^g  -\delta_b^g \delta_c^f) -   \delta_a^f (\delta_b^e \delta_c^g  -\delta_b^g \delta_c^e) + \delta_a^g (\delta_b^e \delta_c^f  -\delta_b^f \delta_c^e),
\eee
which holds in both the Euclidean and Lorentzian context. It implies 
\bee
\epsilon_{abc}\epsilon^{afg}= \delta_b^f \delta_c^g  -\delta_b^g \delta_c^f,
\eee 
which, in turn, is equivalent to 
\bee
\label{special}
[X,[Y,Z]]=\langle X,Z\rangle \,Y -\langle X,Y\rangle \,Z, \quad \forall X,Y,Z\in \cg.
\eee
This  can be used to prove the useful result
\bee
\label{use1}
\langle[X,Y],V\rangle\langle V,Z\rangle  +
\langle [Y,Z],V\rangle \langle V,X \rangle  +
\langle [Z,X],V\rangle \langle V,Y\rangle  =\langle V,V\rangle \langle [X,Y],Z \rangle
\eee
for any $X,Y,Z,V\in \cg$, see \cite{MSkappa} for details and a related identity. 

\subsection{Reformulating the factorisation condition}
\label{tricks}
The Lie bracket or, equivalently, the epsilon tensor provide an identification of $\cg \wedge \cg$ with $\cg$. It follows that for any linear map $\MM:\cg \rightarrow \cg$,  the assignment
\bee
(X,Y)\in \cg \wedge \cg \mapsto [\MM(X),\MM(Y)]\in \cg
\eee
defines a linear map $\cg \rightarrow \cg$. Our factorisation condition \eqref{Mcond} allows for a convenient `dual' formulation   in terms of  this map, which turns out to be the adjugate of $\MM$:
\begin{lemma}
For every linear map $\MM: \cg\rightarrow \cg$, there is a uniquely determined linear map
\bee
\MM^\Ad:\cg \rightarrow \cg, 
\eee
which satisfies
\bee
\label{deftau}
\langle \MM^\Ad (Z),[X,Y]\rangle = \langle Z,[\MM(X),\MM(Y)]\rangle \quad \forall X,Y,Z \in \cg.  
\eee
It is  given by
\bee
\label{TMaster}
\MM^\Ad= \MM^2-\tr(\MM)\, \MM +\frac 1 2 \left (\tr(\MM))^2-\tr(\MM^2)\right)\id ,
\eee
which is the adjugate of $\MM$. 
\end{lemma}
{\bf Proof} \,  Inserting  the basis elements  $J_a,J_b,J_e $ for $X,Y,Z$ in \eqref{deftau} , one deduces the matrix relation
\bee
 (\MM^\Ad)^f_{\;\;e} \epsilon_{fab}=\epsilon_{ecd} \MM^c_{\;\;a}\MM^d_{\;\;b}.
\eee
Multiplying with $\epsilon^{gab}$, summing over $a,b$ and repeatedly applying \eqref{epsid} one arrives at the formula \eqref{TMaster}. To see why \eqref{TMaster} gives the adjugate of $\MM$ (usually defined as the transpose of the matrix of co-factors) note that for any $n\times n$ matrix $A$, the characteristic polynomial $p_A(t) =\det (A-t \,\id)$ has the constant term $(-1)^n\det A$ so that 
\bee
q_A(t)=\frac{p_A(t)-(-1)^n\det A}{t}
\eee
is a polynomial of degree $n-1$ which  satisfies 
\bee
q_A(A) A= A\,q_A(A)= (-1)^{n-1} \det(A)\, \id,
\eee 
by the   Cayley-Hamilton Theorem. 
It is proved in \cite{Taussky} that, in fact,
\bee
A^\Ad=q_A(A).
\eee
The expression \eqref{TMaster} is easily seen to be $q_\MM(\MM)$ for $n=3$
so that the solution of \eqref{deftau} is indeed the adjugate of $\MM$ as claimed.   \hfill $\Box$

Note that for $R\in SO(3)$ or  $R\in SO(2,1)$, the invariance of the epsilon tensor implies $R^\Ad=R^{-1}$.
Inserting this into \eqref{TMaster} is simply the Cayley-Hamilton theorem for the $3\times 3$ matrix $R$.

\begin{lemma}
In the case $\cg=so(3)$ or $\cg= so(2,1)$, the factorisation condition \eqref{Mcond} for the linear map $\MM:\cg\rightarrow \cg$ is equivalent to the quadratic relation
\bee
\label{Mconddd}
(\MM-\tr \MM\, \id )(\MM+\MM^t)+\frac 1 2 \left((\tr \MM)^2-\tr(\MM^2)\right)\id  =-\lambda\id
\eee
\end{lemma}
{\bf Proof}\; 
The factorisation  condition \eqref{Mcond}   can equivalently be written as 
\begin{align}
\label{Mcondd}
\langle Z, [\MM(X),\MM(Y)]- \langle \MM^t (Z),[X,\MM(Y)]+[\MM(X),Y]\rangle \rangle  \qquad \qquad \nonumber \\
= -\lambda\langle Z, [X,Y]\rangle  \quad \forall \; X,Y,Z \in\cg.
\end{align}
Now observe that 
\bee
[X,\MM(Y)]+ [\MM(X),Y]=[(\id +\MM)(X),(\id +\MM)(Y)]-[\MM(X),\MM(Y)]-[X,Y],
\eee
%and hence 
%\begin{align}
%[M(X),M(Y)]-M([X,M(Y)]+ [M(X),Y])  \qquad \qquad  \qquad \qquad  \qquad \qquad  \qquad \qquad \nonumber \\
%=(\id +M)([M(X),M(Y)])+M([X,Y])--[(\id +M)(X),(\id +M)(Y)]
%\end{align}
and apply \eqref{TMaster} both to  $\MM$ and $\id +\MM$ to deduce \eqref{Mconddd}.
\hfill $\Box$

\subsection{Exploiting rotational invariance}
In order to analyse the (equivalent) conditions \eqref{Mcond} and \eqref{Mconddd} further, we 
 split  $\MM$ into symmetric and antisymmetric parts  with respect to $\langle\;\;,\;\;\rangle$. The antisymmetric part can be written as the adjoint action of a general element $V\in\cg$ by virtue of the identity
\bee
\langle [X,Y],Z\rangle + \langle Y,[X,Z]\rangle =0 \quad \forall X,Y,Z \in \cg.
\eee
Thus we write the map $\MM$ as 
\bee
\label{split}
\MM=S + \ad_V.
\eee
for $V\in \cg$ and  $S:\cg\rightarrow \cg$ satisfying
\bee
\langle S(X),Y\rangle = \langle X,S(Y)\rangle  \quad \text{for all} \quad X,Y \in \cg.  
\eee
Inserting  the split \eqref{split}  for $\MM$ and  simplifying  using \eqref{use1} and standard identities we deduce
\begin{align}
\label{maybeuseful}
\langle Z,[S(X),S(Y)]\rangle - \langle X,[S(Y),S(Z)]\rangle - \langle Y,[S(Z),S(X)]\rangle \qquad \qquad \qquad \nonumber \\
+\langle 2 [V,S(Z)]+(\lambda +\langle V,V\rangle) Z ,[X,Y]\rangle =0\quad \forall X,Y,Z \in \cg,
\end{align}
which is equivalent to 
\begin{align}
\label{master}
[S(X),S(Y)]-S([X,S(Y)]+ [S(X),Y]) \qquad \qquad \qquad \qquad \qquad \qquad  \nonumber \\
+ 2(\langle V,X \rangle S(Y) - \langle V,Y\rangle S(X))
=-(\lambda +\langle V,V\rangle)[X,Y] \quad \forall X,Y \in \cg. 
\end{align}
This turns out to be a very useful formulation of the  factorisation condition for the  Lie algebras \eqref{isolie}.

We can derive an equivalent condition to \eqref{master} by  inserting the split \eqref{split} into \eqref{Mconddd}  and using $\MM+\MM^t=2S$. A short calculation gives
\bee
\label{masterr}
2S^2-2\tr(S)\, S+\frac 1 2 ((\tr(S))^2-\tr(S^2))\id + 2  \ad_V S= - (\lambda +\langle V,V\rangle)\id.
\eee
This can also be derived directly from \eqref{master} using the methods of Sect.~\ref{tricks}.
The equation \eqref{masterr} is  invariant under $SO(3)$  or $SO(2,1)$  conjugation, and this can be used to split it into irreducible components under this action, namely into symmetric traceless, antisymmetric and scalar matrices. Noting that the commutator $[\ad_V,S]$
is symmetric but that the anticommutator $\{\ad_V,S\}=\ad_V S+S\, \ad_V $ is antisymmetric we write 
\bee
2  \ad_V S =[\ad_V,S]+\{\ad_V,S\}
\eee
and deduce three equations. For the scalar part we have 
\bee
\label{scalar}
\frac 1 6 \left((\tr S)^2 - \tr(S^2)\right) =\lambda +\langle V, V\rangle,
\eee
for the vector part we find 
\bee
\label{vect}
\{\ad_V,S\} =0
\eee
and for the symmetric, traceless part we deduce
\bee
\label{symtr}
[\ad_V,S]+ 2\left(S^2-\frac 1 3 \tr (S^2)\id \right) -2\left(\tr S\, S -\frac 1 3 (\tr S)^2\id \right)=0.
\eee
The equations \eqref{scalar}, \eqref{vect} and \eqref{symtr} are derived in a different form and by a different method in \cite{Stachura}, where they are   used for  determining the most general $r$-matrix for the Poincar\'e and Euclidean  Lie algebras. 
Adding  \eqref{vect} and \eqref{symtr}  we deduce the  relation
\bee
\label{handy}
 \ad_V S+ \left(S^2-\frac 1 3 \tr (S^2)\id \right) -\left(\tr (S)\, S -\frac 1 3 (\tr S)^2\id \right)=0.
\eee
This has a various useful consequences, including the following lemma which is also implicit in some of the manipulations in the appendix of \cite{Stachura}.
\begin{lemma}
\label{neat}
If the map $\MM:\cg\rightarrow \cg$ solves the factorisation condition \eqref{Mconddd} and $S$ and $\ad_V$ are its symmetric and antisymmetric part as in \eqref{split}, then the following hold:
\begin{enumerate}
\item[(1)] The antisymmetric part $\ad_V$ is the zero map on any subspace of $\ch\subset \cg$ where the restriction $S_{|\ch}$ is invertible. In particular, if $S$ is invertible then $V$ vanishes.  
\item[(2)] If  dim ker $S$=1 then: 
 ker $S$ is not a null space  $\Rightarrow$  $V = 0$. 
\end{enumerate}
\end{lemma}
In the following, the contrapositive of result (2) will be particularly useful: if  ker $S$ is one-dimensional and $V\neq 0$ then  ker $S$ is necessarily a null space.

\noindent {\bf Proof} \;  (1)  On any subspace $\ch$ where $S_{\ch}$ is invertible we can multiply \eqref{handy}  by the (symmetric) map $S_{|\ch}^{-1}$ to obtain an expression for 
the antisymmetric map $\ad_V$ in terms of a symmetric map. Thus $(\ad_V)|_{|\ch}=0$.

\noindent (2) \, It follows from \eqref{vect} that ker $S$ is invariant under the action of $\ad_V$. If ker $S$ is one-dimensional then any basis vector $X$ of it is necessarily an eigenvector of $\ad_V$, i.e., $[V,X]=\mu X$. Since $\langle [V,X], X\rangle =0$, we deduce that $\mu=0$ if $X$ is not null. However, in that case, $S$ restricted to the orthogonal complement of $X$ is invertible, so we  know from  part (1) that $\ad_V$ vanishes on the orthogonal complement of $X$. Since we already have $\ad_V(X)=0$ we deduce that $V=0$.   \hfill $\Box$

\section{Solving the factorisation condition}

\subsection{Solutions of definite symmetry}
Before we systematically study solution of the equation \eqref{master} in the next sections, we note some special cases which can be found by inspection.
Setting the symmetric part $S$ to zero, the condition \eqref{master} reduces to a condition on the Lie algebra element $V$. Expanding
\bee
\label{vexp}
V=-v^aJ_a,
\eee
where the sign is chosen to match conventions in \cite{OS1}, we have 
\bee
\label{vcond}
\langle V, V\rangle = v^a v_a =-\lambda.  
\eee
The resulting bialgebra structure on the semidual Lie algebra  has the $r$-matrix
\bee
\label{bicr}
r_\kappa=v^c \epsilon^{b}_{\;\; ac} \,\cP^a \wedge J_b, 
\eee
which is the familiar 3d $\kappa$-Poincar\'e structure with deformation parameter $\bv=(v^0,v^1,v^2)$, which may be timelike, spacelike or timelike, depending on $\lambda$ \cite{BHDS,BHDSspacelikedeformation,BHDSlightlikedeformation}. The resulting Lie algebra structure of $\cm$ is a semidirect sum $\RR \ltimes \RR^2$, with $\RR$ acting on  $\RR^2$ by scaling.  We will recover this solution as part of a more general family in the next section.

Similarly, setting the antisymmetric part $\ad_V$ of the map $\MM$ to zero,  we 
 deduce from equations \eqref{symtr} and \eqref{scalar} that 
\bee
\label{vzero}
 S^2-\tr(S) \, S =- 2\lambda\, \id \quad \text{for} \quad  V=0
\eee
At the same time, we know from the Cayley Hamilton theorem that 
\bee
S^3-\tr(S) \, S^2 + \frac 1 2 \left((\tr(S))^2-\tr(S^2)\right)S-\det(S)\id =0.
\eee
Thus,  from \eqref{scalar} with $V=0$
\bee
(S^2-\tr(S) \, S + 3\lambda\id  )S-\det(S)\id =0.
\eee
Combining with \eqref{vzero} we deduce
\bee
\lambda S =\det S \; \id \quad  \text{for} \; V=0.
\eee
It follows that for $\lambda >0$ any purely symmetric solutions is diagonal and given by
\bee
\label{diagsol}
\MM=\sqrt{\lambda} \id.
\eee 
The resulting classical $r$-matrix
\bee
\label{doubler}
r_{\text{\tiny double}}= \sqrt{\lambda} \,\cP^a \wedge J_a
\eee
is that of the classical double of $so(3)$ or $so(2,1)$. The associated Lie brackets on $\cm$ are 
\bee
[\Pp_a,\Pp_b]=2\sqrt{\lambda}\epsilon_{abc}\Pp^c,
\eee
i.e., $\cm$ is $so(3)$ or $so(2,1)$ in this case.

When $\lambda = 0$ we have another symmetric solution that can be found by inspection (but which we will recover systematically later).  For any vector $\bm$,  the liner map with matrix
\bee
\label{trivsol}
\MM^b_{\;\;a} = m^bm_a.
\eee
trivially satisfies \eqref{master} (for $\lambda =0$), so that 
\bee r_m=m_a m^b  \cP^a \wedge J_b
\eee is a classical $r$-matrix for the Euclidean or the Poincar\'e Lie algebras.
The Lie algebra structure of $\cm$ turns out to be a semidirect sum $\RR\ltimes \RR^2$ with $\RR$ acting on $\RR^2$ by rotations, Lorentz boosts or the nilpotent matrix 
\bee
\label{jord}
M =\begin{pmatrix} 0 & 1 \\ 0 & 0 \end{pmatrix}, 
\eee
depending on whether $\bm$ is timelike, spacelike or lightlike. We will study this solution, too, as part of a larger family in the next sections and give more details there.

\subsection{The standard case}

In the simplest case, the symmetric part $S$ has a diagonalising orthonormal basis $K_a$, $a=0,1,2$,  of $\cg$ with real eigenvalues $\lambda_a$, $a=0,1,2$. The diagonalisation is always possible in the Euclidean case but cannot always be achieved in the Lorentzian case. We will consider the other, non-standard cases below.
Since the  diagonalising basis is related to the original basis $J_a$, $a=0,1,2$ by an orthogonal (i.e., $SO(3)$ or $SO(2,1)$)  transformation, we know that   the commutators are still 
\bee
[K_a,K_b]=\epsilon_{abc} K^c.
\eee
The map $S$  in \eqref{split} then takes the from 
\bee
\label{ansatz}
S=\sum_{a=0}^2 \lambda_a \ket{K_a}\bra{K_a}.
\eee
In order to find all solutions of this form, we consider two cases.\\
\noindent (I)  dim ker$S \leq 1$. In that case, it follows  from Lemma \ref{neat} that $V=0$: if $S$ is invertible this is a consequence of part (1) while for a one-dimensional kernel of $S$ the vanishing of $S$ follows from part (2) since  ker $S$ cannot be a null space in the diagonalisable case under consideration. With $V=0$ we insert the diagonal form of $S$   \eqref{ansatz}  into  \eqref{vzero} to find
\begin{align}
\lambda_0(\lambda_1+\lambda_2)&=2\lambda, \nonumber \\
\lambda_1(\lambda_0+\lambda_2)&=2\lambda, \nonumber \\
\lambda_2(\lambda_0+\lambda_1)&=2\lambda.
\end{align}
Taking pairwise differences we deduce 
\bee
\lambda_0=\lambda_1=\lambda_2=\sqrt{\lambda}.
\eee
This is the solution \eqref{diagsol} already found by inspection. 
Since the eigenvalues $\lambda_a$ are assumed to be real,  this is only a valid and  non-trivial solution if  $\lambda >0$.
 
\noindent (II)  dim ker $S \geq 2$. In that case,  $S^2=\tr(S) S$, so that the scalar equation \eqref{scalar} gives the condition \bee
\label{vcondd}
\langle V, V\rangle = -\lambda
\eee
  for $V,$ and the matrix condition \eqref{symtr}  reduces to 
\bee
[\ad_V,S]=0. 
\eee
If $S=0$ then this imposes no further restriction of $V$ and we recover the purely antisymmetric  solution  $\MM=\ad_V$ found by inspection in the previous section. However, if dim ker $S$=2 so that $S=\lambda_a \ket{K_a}\bra{K_a}$ for some fixed $a$ and $\lambda_a\neq 0$, we see that we can now choose $V$ to be a multiple of $K_a$, with the multiple chosen so  that the normalisation condition \eqref{vcond} holds. When $\lambda \neq 0$, the resulting solution can be written
\bee
\label{firstnew}
\MM=\beta  \ket{V}\bra{V}  + \ad_V, 
\eee
with $\beta \in \RR$ arbitrary and $V$ satisfying \eqref{vcondd}.  When $\lambda =0$, the requirement that $S=\lambda_a \ket{K_a}\bra{K_a}$ (no sum) and $\ad_V$ commute enforces  $V=0$.  In that case we recover the solution \eqref{trivsol}, at least for space- or timelike vectors $\bm$. The case of light-like $\bm$ represents a non-diagonalisable map $S$ and will appear in Sect.~\ref{nonstandard}.

In the remainder of this subsection we study the solution  \eqref{firstnew} in more detail. With the convention \eqref{vexp}, the generators of the subalgebra $\cm$ in this case are 
\bee
\label{newgen}
\Pp_a=\P_a +\epsilon_{abc}v^bJ^c + \beta(v^bJ_b) \, v_a,
\eee
where $\beta$ is an arbitrary (real) parameter. 
Using 3-vector notation $\bv=(v^0,v^1,v^2)$ etc.  we can also write
\bee
\label{newgenn}
\bPp=\bP +\bv \times \bJ + \beta (\bv\cd \bJ)  \, \bv. 
\eee
Then, with $\bv\cd\bw=v^a w_a$ etc.,  the brackets  are 
\begin{eqnarray}
\label{sublieP}
\!\!\!\!\!\!\!\!\!\!\!\!\!\!\!\!\!\![\bp\cd \bPp,\bq\cd\bPp]&\!\!\!=\!\!\!\!&-\bv\times(\bp\times \bq)\cd(\bPp-\beta \bv\times \bPp) \nonumber \\
&\!\!\!=\!\!\!\!&(\bv\cd\bp)(\bq\cd\bPp)-(\bv\cd\bq)(\bp\cd\bPp) -\beta (\bp\times\bq)\cd\bv \; (\bv\cd\bPp) - \lambda\beta(\bp\times\bq)\cd\bPp
\end{eqnarray}
as well as 
\bee
\label{subliePJ}
[\bp\cd \bJ,\bq\cd\bPp]=\bp\times \bq\cd\bP+\bp\times (\bq\times \bv)\cd \bJ + \beta (\bq\cd\bv)\, \bp\times \bv \cd \bJ,
\eee
which can be written in terms of the generators $\Pp_a$ as 
\begin{align}
\label{subliePpJ} 
[\bp\cd \bJ,\bq\cd\bPp]& =\bp\times \bq\cd\bPp+ (\bv\times \bp -\beta \bv \times (\bv \times \bp))\cd \bq\times \bJ. \nonumber \\
& =\bp\times \bq\cd\bPp +(\bv\cd \bq)(\bp\cd\bJ) -(\bp\cd\bq)(\bv\cd \bJ) -\beta(\bp\cd\bv)(\bv\times\bq\cd \bJ)- \lambda\beta (\bp\times\bq\cd \bJ) .
\end{align}
To identify the resulting Lie algebra structure of $\cm$, we consider  the various cases. 

\subsubsection{The Euclidean case with $\lambda <0$} 
We pick $
\bv =\sqrt{-\lambda}(1,0,0)$, 
so that  \eqref{newgenn} gives
\bee
\Pp_0 =\P_0-\beta\lambda J_0,\quad \Pp_1=\P_1-\sqrt{-\lambda}J_2,\quad \Pp_2=\P_2+\sqrt{-\lambda}J_1.
\eee
Then \eqref{sublieP} yields the following brackets for $\cm$:
\begin{align}
 [\Pp_1,\Pp_2]&=0,\nonumber \\
[\Pp_0,\Pp_1]&=\sqrt{-\lambda}\Pp_1-\beta \lambda \Pp_2,\nonumber \\
[\Pp_0,\Pp_2]&=\sqrt{-\lambda}\Pp_2+\beta \lambda \Pp_1.
\end{align}
Thus $\Pp_1$ and $\Pp_2$ span a commutative subalgebra, and $\Pp_0$ acts on this by infinitesimal scaling and rotation. In terms of the Bianchi classification of three-dimensional real Lie algebras, this is type VII.

The commutators  involving $J_a$ and $\Pp_b$ are 
\begin{align}
 [J_0,\Pp_1]&= \Pp_2, \quad  [J_0,\Pp_2]  = -\Pp_1,\nonumber \\
[J_1,\Pp_0]&= \sqrt{-\lambda} J_1+\beta \lambda J_2  -\Pp_2, \quad [J_2,\Pp_0]  = \sqrt{-\lambda} J_2-\beta \lambda J_1  + \Pp_1, \nonumber \\
[J_1,\Pp_2]&= \Pp_0- \beta \lambda J_0, \quad [J_2,\Pp_1]  = - \Pp_0+\beta \lambda J_0,  \nonumber \\
[J_0,\Pp_0]&= 0, \quad [J_1,\Pp_1] =[J_2,\Pp_2]= \sqrt{-\lambda}J_0,
\end{align}
and again allow for a geometric interpretation. The action of $J_a$ on  the vector $\bPp$ is infinitesimal rotation around the $a$-th axis.  The action of $\Pp_0$ on the $J_1J_2$-plane  is  infinitesimal scaling and rotation, as on the $\Pp_1\Pp_2$-plane above.  The action of both $\Pp_1$ and $\Pp_2$  on  the $J_1J_2$-plane is to map it onto $J_0$, which is consistent with the fact that $\Pp_1$ and $\Pp_2$ commute.

\subsubsection{Lorentzian case with $\lambda<0$}

If $\bv $ is the timelike vector $
\bv  =\sqrt{-\lambda}(1,0,0)$, 
then  the generators (\ref{newgen})  of $\cm$ are 
 \bee
\Pp_0 =\P_0-\beta\lambda J_0,\quad \Pp_1=\P_1+\sqrt{-\lambda}J_2,\quad \Pp_2=\P_2-\sqrt{-\lambda}J_1,
\eee
so that from (\ref{sublieP}) we obtain 
\bea
\left[\Pp_1,\Pp_2\right]&=&0,\nonumber\\
\left[\Pp_0,\Pp_1\right]&=&\sqrt{-\lambda}\Pp_1+\beta\lambda \Pp_2, \nonumber \\
\left[\Pp_0,\Pp_2\right]&=&\sqrt{-\lambda}\Pp_2-\beta\lambda \Pp_1. 
\eea
Again we have a two-dimensional  abelian algebra  spanned by $\Pp_1, \Pp_2$, with $\Pp_0$ acting by infinitesimal rotation and scaling. The Bianchi type is again  VII. 
Computing the mixed commutators  is straight forward and leads to a geometric interpretation analogous to the Euclidean case, but we omit the details here.

\subsubsection{Lorentzian case with $\lambda>0$}
With the spacelike vector  $
\bv  =\sqrt{\lambda}(0,1,0)$, 
we have from (\ref{newgen}) that
\bee
\Pp_0 =\P_0-\sqrt{\lambda} J_2,\quad \Pp_1=\P_1-\beta \lambda J_1,\quad \Pp_2=\P_2-\sqrt{\lambda}J_0,
\eee
and therefore  (\ref{sublieP}) gives
\bea
 \label{spacelikePP}
\left[\Pp_0,\Pp_2\right]&=& 0,\nonumber\\
\left[\Pp_1,\Pp_0\right]&=& -\sqrt{\lambda}\Pp_0-\beta\lambda \Pp_2 , \nonumber \\
\left[\Pp_1,\Pp_2\right]&=& -\sqrt{\lambda}\Pp_2-\beta\lambda \Pp_0 .
\eea
It is illuminating to write the commutators in this case in terms of the generators $N,\tilde N,J_1$ defined in \eqref{newgens} and the corresponding  generators
\bee
\P_N=\frac{1}{\sqrt{2}}(\P_0+\P_2),\quad \P_{\tilde{N}}=\frac{1}{\sqrt{2}}(\P_0-\P_2).
\eee
Then 
\bee
\Pp_N=\P_N-\sqrt{\lambda}N,\quad \Pp_{\tilde{N}}=\P_{\tilde{N}}+\sqrt{\lambda}\tilde{N}, \quad \Pp_1=\P_1-\beta \lambda J_1,
\eee
and therefore  (\ref{spacelikePP}) gives
\bea 
\label{spacelikePPN}
\left[\Pp_N,\Pp_{\tilde{N}}\right]&=&0,\nonumber\\
\left[\Pp_1,\Pp_N\right]&=&-(\beta\lambda+\sqrt{\lambda})\Pp_N, \nonumber \\
\left[\Pp_1,\Pp_{\tilde{N}}\right]&=&(\beta\lambda-\sqrt{\lambda})\Pp_{\tilde{N}},
\eea
This is again a semidirect sum $\RR\ltimes \RR^2$ and Bianchi type VI: $\Pp_N$ and $\Pp_{\tilde N}$ span a two-dimensional abelian algebra, and $\Pp_1$ acts on them with a matrix that has (generically distinct) real eigenvalues. 

\subsection{The non-standard case}
\label{nonstandard}
In the Lorentzian case,  maps $S:\cg\rightarrow \cg$  which are symmetric with respect to the non-degenerate symmetric form $\langle \cdot ,  \cdot \rangle$ cannot always be diagonalised. We now consider the different normal forms (which are discussed, for example, in the textbook \cite{Hall}). In each case,  the solution of the factorisation condition \eqref{master} or, equivalently,  Eq.~\eqref{masterr} uses Lemma \ref{neat} and standard arguments from linear algebra; some proceed along similar lines to the discussion in the appendix of \cite{Stachura}. 

\subsubsection{Small Jordan Block}
In this case the  symmetric map $S$ has a lightlike and a spacelike eigenvector and can be brought into the form 
\bee
S=a (N \bra{\tilde N} +  \tilde N \bra{N}) + b N\bra{N} - c J_1\bra{J_1}, \qquad b\neq 0, 
\eee
so that 
\bee
 S(N)= a N,   \quad 
S(\tilde N)=   bN +a \tilde N,    \quad
S(J_1)= cJ_1.
\eee
It is easy to check that, if $a\neq 0$ and $c\neq 0$ (so that $S$ is invertible) we have $V=0$ by Lemma \ref{neat} and hence the diagonal solution \eqref{diagsol}. If $a\neq 0$  then $S$ has a one-dimensional kernel which is not null, so by Lemma \ref{neat} we again deduce $V=0$ and obtain a  solution of the form \eqref{trivsol}.  If $a=0$ but $c\neq 0$ we have, from \eqref{scalar}, that $\langle V, V\rangle =-\lambda$. The requirement that $\ad_V$ leave the kernel of $S$ invariant implies that $V$ is spacelike or lightlike, i.e.,  $\lambda \geq 0$. When $\lambda >0$, we can use \eqref{handy} to deduce that $c=\sqrt{\lambda}$. Renaming $b$ as $\beta$ we 
have  the new solution
\bee
\label{smalljordan}
\MM= \beta \ket{N}\bra{N} -\sqrt{\lambda}\ket{J_1}\bra{J_1} +\sqrt{\lambda}\;\ad_{J_1}, \quad \lambda >0, \quad \beta \in \RR,
\eee
so that 
\bee
\Pp_N=
\P_N+\sqrt{\lambda}N,\quad \Pp_{\tilde{N}}=\P_{\tilde{N}}+\beta N-\sqrt{\lambda}\tilde{N}, \quad \Pp_1=\P_1+\sqrt{\lambda}J_1.
\eee
The commutators of $\cm$ are
\bea \label{spacelikePPNb}
\left[\Pp_N,\Pp_{\tilde{N}}\right]&=&0,\nonumber\\
\left[\Pp_1,\Pp_N\right]&=&2 \sqrt{\lambda}\Pp_N, \nonumber \\
\left[\Pp_1,\beta \Pp_N-2 \sqrt{\lambda}\Pp_{\tilde{N}}\right]&=&0.
\eea
This is Bianchi type III, which is  isomorphic to  the direct sum of the one dimensional Lie algebra $\RR$  (generated here by $ \beta  \Pp_N -2 \sqrt{\lambda} \Pp_{\tilde{N}}$ ) and  the  non-abelian two-dimensional Lie algebra $L(2)$ (generated  here by  $\Pp_N,\Pp_{\tilde{N}}$).

When $\lambda =0$,  we again use \eqref{handy} to obtain the solution
\bee
\label{lightjordan}
\MM=\beta \ket{N}\bra{N} +\ad_N,
\eee
for an arbitrary $\beta \in \RR$. This may be viewed as a limit as $\lambda \rightarrow 0$ of  \eqref{smalljordan} and is also a  lightlike version of \eqref{firstnew}. The generators of $\cm$ are 
\bee
\Pp_N=\P_N,\quad \Pp_{\tilde{N}}=\P_{\tilde{N}}-J_1+\beta N, \quad \Pp_1=\P_1-N,
\eee
and the commutators of $\cm$  are 
\bea 
\left[\Pp_N,\Pp_1\right]&=&0\nonumber \\
\left[\Pp_{\tilde{N}},\Pp_N\right]&=&-\Pp_N\nonumber\\
\left[\Pp_{\tilde{N}},\Pp_1\right]&=&-\Pp_1-\beta \Pp_N. 
\eea
which are again those of a semidirect sum  $\RR\ltimes \RR^2$. When $\beta =0$,  this is Bianchi type V, with   $- \Pp_{\tilde N}$ acting on the span of $\Pp_N,\Pp_1$ via the identity. When $\beta \neq 0$, it is Bianchi type IV, with 
 $-\frac{1}{\beta} \Pp_{\tilde N}$ acting  on  the span of $\Pp_N,\Pp_1$ via $\id/\beta  + M$, where $M$ is the nilpotent linear map with  matrix \eqref{jord}.

\subsubsection{Large Jordan Block}
The symmetric map $S$ only has a single, lightlike eigenvector and can be brought into the form
\bee
S=  a (N \bra{\tilde N} +  \tilde N \bra{N}-  J_1\bra{J_1}) + b (N \bra{J_1} +  J_1 \bra{ N}), \qquad b\neq 0
\eee
so that 
\bee
S(N)= a N,  \quad 
S(\tilde N)= a \tilde N   + b J_1,  \quad
S(J_1)= - bN  + a J_1. 
\eee
Again we require $a=0$ for a non-vanishing $V$. Then  ker $S$ is one-dimensional and  spanned by the lightlike generator $N$; this space  is  invariant under $\ad_V$ provided  $V$ is a linear combination of   $N$ and $J_1$.  However, since $\tr S=\tr S^2=0$ we can use \eqref{handy} to  deduce that $\ad_V S + S^2=0$, which in turn implies that 
$V=-b N$.
Thus we are in the case  $\lambda=0$ and, after re-naming $b=\beta/2$, we  have the solution
\bee
\MM=\frac{\beta}{2}\left( \ket{N}\bra{J_1}+ \ket{J_1}\bra{N} -\ad_N\right), 
\eee
which can be written more compactly as 
\bee
\label{largejordan}
\MM=\beta \ket{J_1}\bra{N}.
\eee
The resulting basis of $\cm$ is  
\bea
\Pp_N=P_N, \quad \Pp_{\tilde{N}}= \Pp_{\tilde{N}} +\beta J_1, \quad \Pp_1=\P_1,
\eea
and we obtain the commutators
\bee
[\Pp_1,\Pp_N]=0, \quad [\Pp_1,\Pp_{\tilde{N}}]=0, \quad [\Pp_{\tilde{N}},\Pp_N]=\beta \Pp_N.
\eee
This is again a direct sum of $\RR$ (spanned by $\Pp_1$) and the two-dimensional non-abelian Lie algebra, i.e., Bianchi type III

The remaining normal form of a symmetric map $S:\cg\rightarrow \cg$ is 
\bee
S=a (N \bra{\tilde N} +  \tilde N \bra{N}) + b (\tilde N \bra{\tilde N}-  N\bra{N}) - c J_1\bra{J_1}
\eee
which has the form of a `rotation'  in the span of $\{N,\tilde{N}\}$:
\bee
 S(N)= a N + b \tilde N, \quad 
S(\tilde N)= -b  N +  a \tilde N \quad
S(J_1)= c J_1. 
\eee
In this case we do not obtain a new solution: the cases  dim ker $S \leq 1$ can be dealt with as above. 
However,  dim ker $S$=2 requires $a=b=0$, in which case we recover a solution in the family \eqref{firstnew}

\section{Summary and Discussion}
We can unify a large family of solutions of the factorisation condition \eqref{Mconddd}   in the form
\bee 
\label{prettygeneral}
F= \beta V\bra{V} + \alpha\, \ad_V,  \qquad \beta \in \RR,\;\;\; \alpha \in \{0,1\}, \;\;\;\alpha \langle V,V\rangle = -\lambda. 
\eee
This reduces to  the purely symmetric solutions \eqref{trivsol}  when $\alpha =0$ and to the family \eqref{firstnew}, including the lightlike case \eqref{lightjordan}, when $\alpha=1$. This factorisation map gives rise to the $r$-matrices 
\bee
\label{kappagen}
\tilde{r}_\kappa = (\beta v^a v_b + \alpha v^c\epsilon^b_{\;\;ac})P^a\wedge J_b,
\eee
which generalise the familiar $\kappa$-Poincar\'e  $r$-matrix \eqref{bicr};  we therefore call them 
generalised $\kappa$-Poincar\'e solutions.

\begin{figure}[ht]
\centering
\includegraphics[width=3.5truecm]{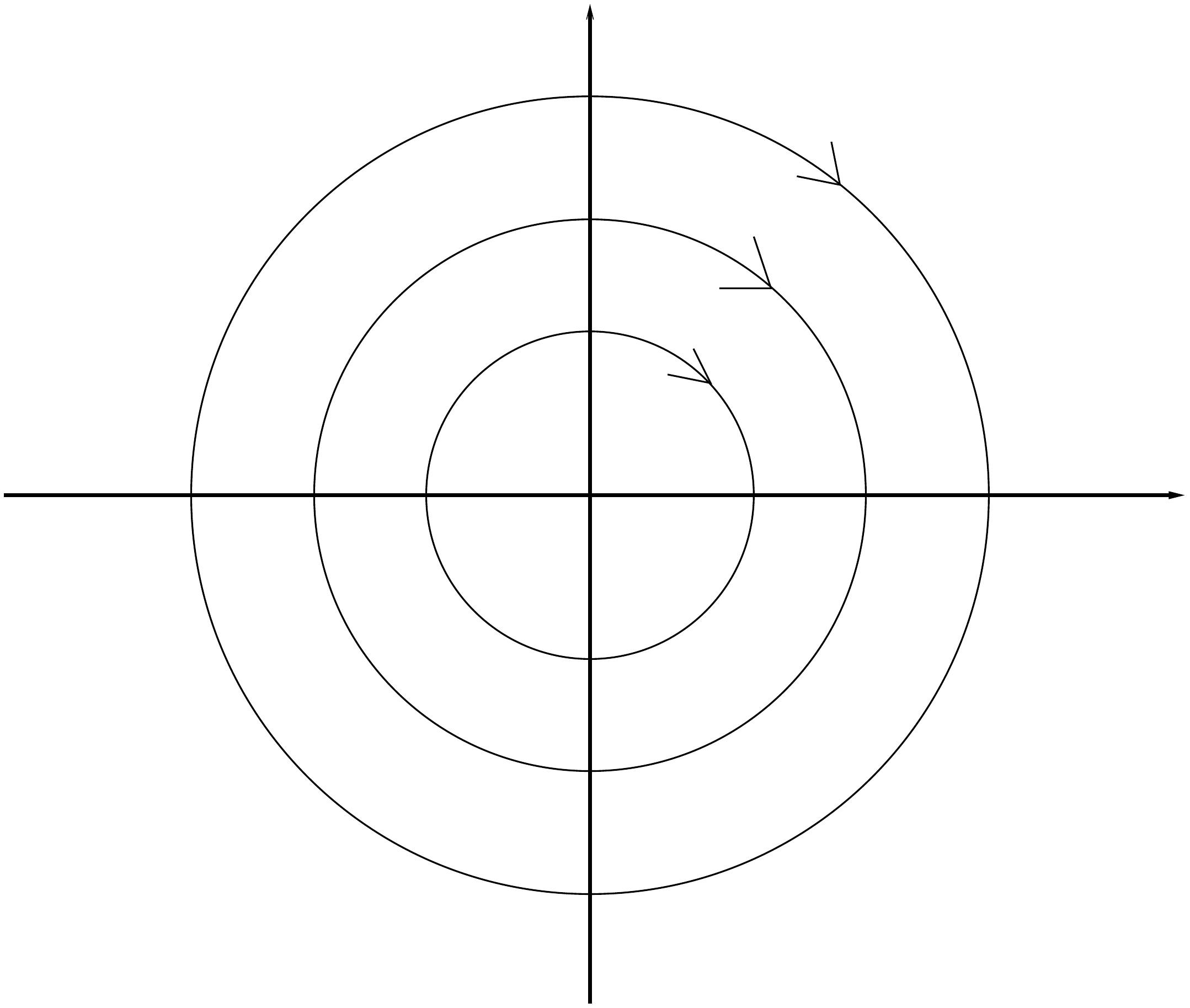}
\includegraphics[width=3.5truecm]{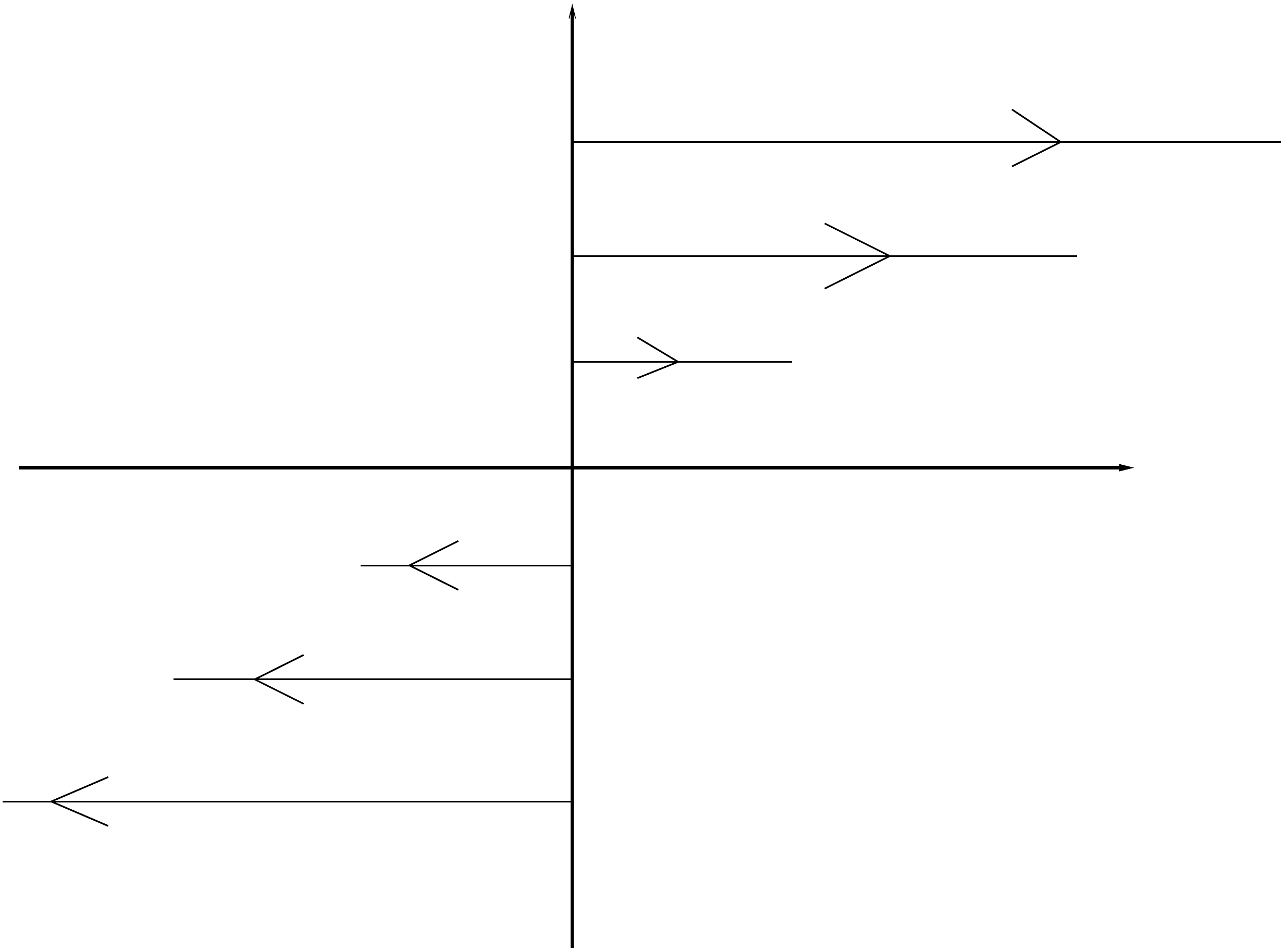}
\includegraphics[width=3.5truecm]{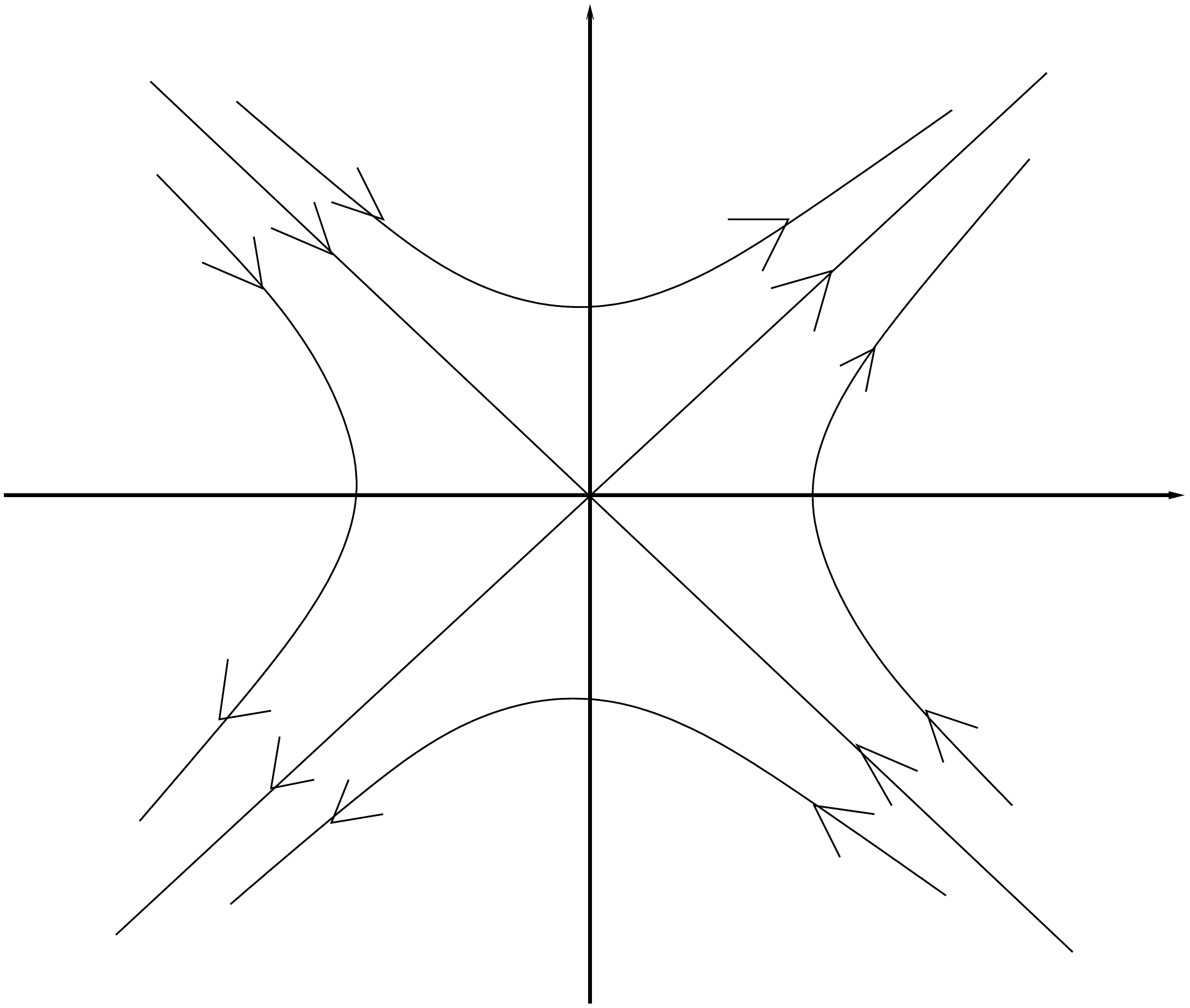}\\
\includegraphics[width=3.5truecm]{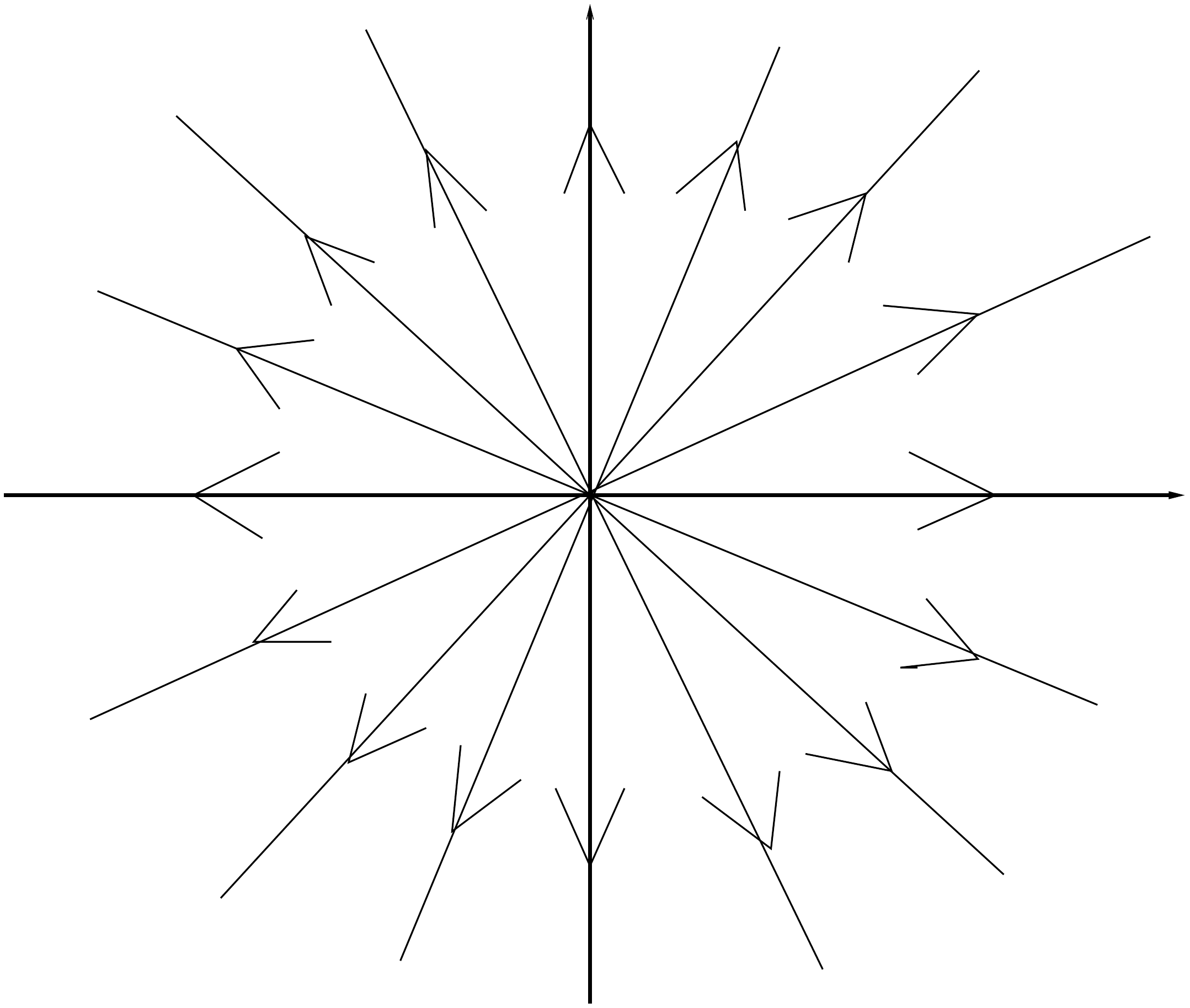}
\includegraphics[width=3.5truecm]{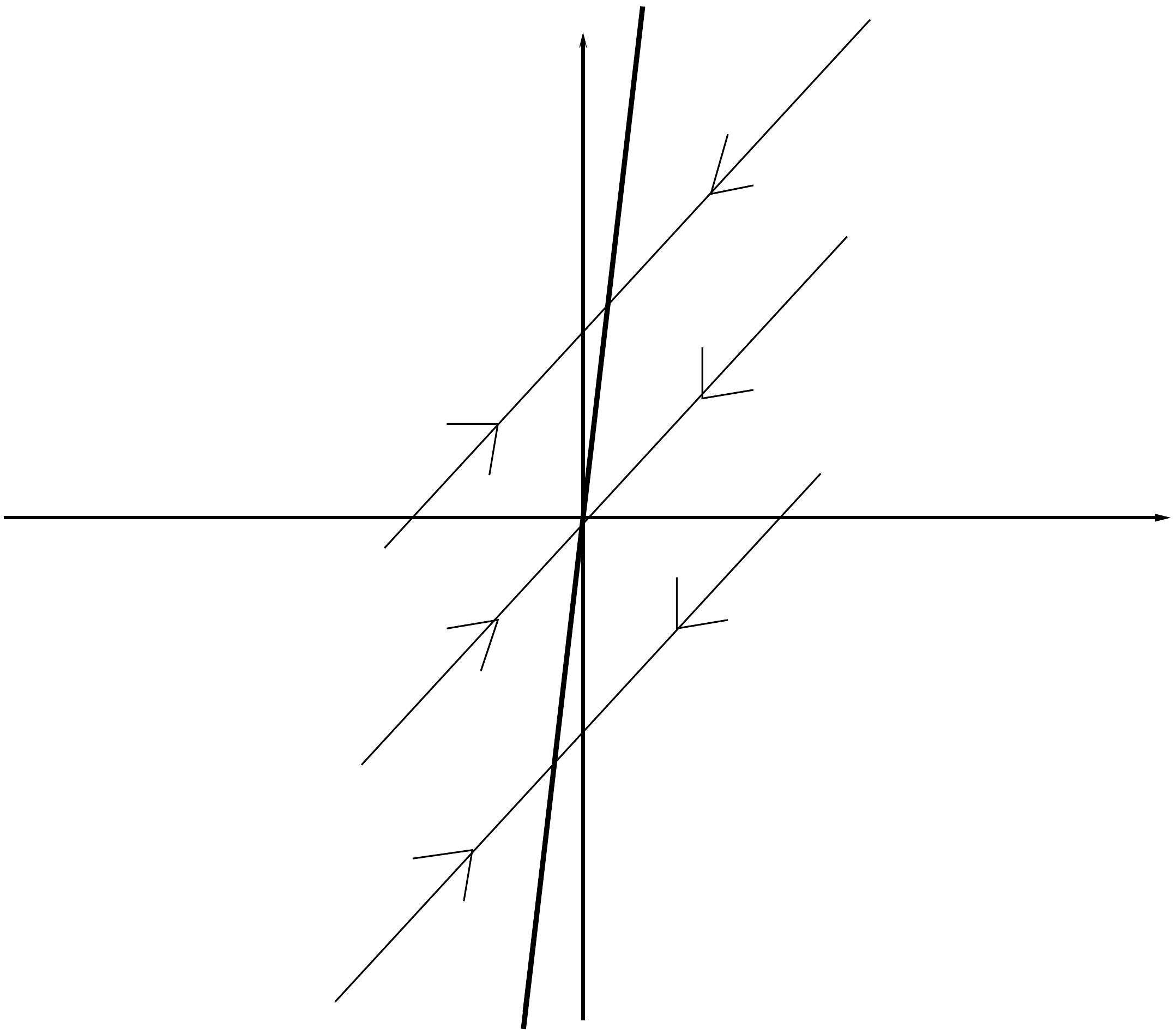}
\caption{The  $\RR$-actions in the semidirect sums $\RR\ltimes \RR^2$  and their degenerate limit $\RR\oplus L(2)$; explanation in the main text}
\label{flowpics}
\end{figure}

The Lie algebra structure of $\cm$ arising from the family \eqref{prettygeneral} is of the semidirect sum form $\RR\ltimes \RR^2$ for all values of the parameters. The $\RR$-action on $\RR^2$ has two components, corresponding to the parameters $\alpha$ and $\beta$ being non-zero.  The $\alpha$-component (for $\alpha =1$ and $\beta=0$) is simple overall scaling,  shown  in the bottom left of  Fig.~\ref{flowpics} and associated with  the standard  $\kappa$-Poincar\'e algebra. The $\beta$-component (for $\beta\neq 0$ and $\alpha =0$) is more interesting and depends on the value of $\lambda$. Its matrix  is proportional to one of the $2\times 2$ matrices  $\rho(\theta)$ representing the relation $\theta^2=\lambda$:
\bee
\rho(\theta) =  
\begin{cases} 
 \begin{pmatrix} 0 &  \sqrt{-\lambda}  \\ -\sqrt{-\lambda} & 0\end{pmatrix} \quad \text{if} \; \lambda <0  \\
\\
 \begin{pmatrix} \phantom{-} 0  & \phantom{-} 1\phantom{-}  \\ \phantom{-} 0 & \phantom{-} 0 \phantom{-} \end{pmatrix} \quad  \quad \; \;\;\text{if} \; \lambda =0 \\
 \\
\begin{pmatrix} \sqrt{\lambda}  & 0 \\ 0 &-\sqrt{\lambda} \end{pmatrix} \quad \quad \;\text{if} \; \;\lambda >0
 \end{cases}
\eee
The flows generated by exponentiating these matrices are shown from left to right  in the top row of Fig.~\ref{flowpics}.

In addition to the family \eqref{prettygeneral} we found the solution $F=\sqrt{\lambda}$ id,  which gives rise  to the $r$-matrix of the classical double \eqref{doubler} and $\cm=so(3)$ or $\cm=so(2,1)$. Finally, we have the exceptional,  Jordan-type solutions \eqref{smalljordan} and  \eqref{largejordan}. With the notation $
P_N=(P_0+P_2)/\sqrt{2}$, 
the $r$-matrix associated to the  factorisation map \eqref{smalljordan} is 
\bee
\label{smallr}
r_{\text{\tiny SJ}}= \beta P_N\wedge N + \sqrt{\lambda}(P_1\wedge J_1 + \epsilon^b_{\;\;a1} P^a\wedge J_b),
\eee
while the $r$-matrix associated to the solution \eqref{largejordan} is simply
\bee
\label{larger}
r_{\text{\tiny LJ}} = \beta P_N\wedge J_1.
\eee
The associated Lie algebra structure of  $\cm$ is $\RR\oplus  L(2)$, which may be viewed as a degenerate case  of the  semidirect sums $\RR\ltimes \RR^2$ with a diagonalisable $\RR$-action  but one eigenvalue vanishing.  The eigenvector for the non-vanishing eigenvalue is always the lightlike vector $N$, but the eigenvector for the zero eigenvalue varies  between the two cases and as a function of  the parameter $\beta$. The flow obtained by exponentiating such matrices is shown for a generic case in  the bottom  right  of Fig.~\ref{flowpics}.

\begin{table}[h]
\centering
\begin{tabular}{|c|c|c|c|}
\hline 
&& &\\
$\MM$ & $r$-matrix & $\cm $& Bianchi type \\
&&&\\
\hline
&&&\\
0 & 0 & $\RR^3$& I \\
&&&\\
 $\sqrt{\lambda} \,\id $ & $r_{\text{\tiny double}}$ & $so(2,1)$ or $ so(3)$ & VIII or IX\\
 &&&\\
$ \beta \ket{V}\bra{V} + \alpha \,\ad_V$ & $\tilde{r}_\kappa$& $\RR\ltimes \RR^2$ & IV-VII\\
 $\beta \in \RR, \alpha \in \{0,1\}$ & & & \\
&&&\\
Jordan forms & $r_{\text{\tiny SJ}}$ and  $r_{\text{\tiny LJ}} $ & $\RR\oplus L(2)$ & III \\
\eqref{smalljordan} and \eqref{largejordan} &  && \\
\hline
\end{tabular}
 \vspace{.5cm}
\caption{Factorisations and associated $r$-matrices}
\label{summary}
 \end{table}

Table~\ref{summary} summarises our results and discussion  of the factorisation of the Lie algebras \eqref{isolie}  and  the associated $r$-matrices of the semiduals \eqref{eucllorentz}.  As reviewed in the Introduction,  semidualisation gives all $r$-matrices of the form $R^b_{\;\;a}P^a\wedge J_b$ in this case. It is interesting that for each  type of $r$-matrices there is an associated Bianchi type for the Lie algebra $\cm$.  All Bianchi types except type II (the Heisenberg algebra) arise in this way. 

Viewing the generators $\Pp_a$ of the original algebra \eqref{EuP} as spacetime coordinates and the  dual generators $P_a$ in \eqref{eucllorentz} as momenta, the  summary in  Table~\ref{summary} may thus also be viewed as a list of non-commutative spaces and associated non-co-commutative   momentum spaces,  parametrised by the relevant $r$-matrices. We thus obtain the  unified picture of non-commutative geometries and isometry algebras promised in the Introduction. 

There are several questions and topics for further research which follow from the results reported here. Mathematically, one would like to understand more generally when semidualising   families  of Lie algebras with  a given double cross sum decomposition provides an effective way of finding $r$-matrices for the semidual Lie bialgebra. It would also be interesting to apply the same technique to families of Lie bialgebras with non-trivial co-commutators, so that the semidual has a more complicated Lie algebra structure than the semidirect sums found here.  

 In the physics literature,  the standard $\kappa$-Poincar\'e (with a timelike deformation parameter) and the classical double   Lie bialgebras in Table~\ref{summary}  have been much studied, the latter (but not the former - see \cite{MSkappa})  being related to 3d quantum gravity.  It would be interesting to see if  the  generalised $\kappa$-Poincar\'e $r$-matrices \eqref{kappagen}  with $\beta\neq 0$ and  the Jordan-type $r$-matrices 
\eqref{smallr} and \eqref{larger}  also   have applications in real or toy models of mathematical physics. 

\noindent {\bf Acknowledgments}  Some of the work reported in this paper was carried out while Prince K Osei was supported by an ICTP PhD fellowship. We thank the Perimeter Institute for hospitality and support  during the final stage of the project.


\begin{thebibliography}{99}


\bibitem{Majidbicross}
S.~Majid, Physics for algebraists: noncommutative and noncocommutative Hopf algebras by a bicrossproduct construction, J.~Algebra, 130 (1990) 17--64. 
%still Hopf algebras

 \bibitem{Majid90} S. Majid, Matched pairs of Lie groups associated
to solutions of the Yang-Baxter equations, Pac.~J.~Math.  141 (1990)  311--332.
%Lie groups



\bibitem{Majid} S.~Majid, Foundations of quantum group theory,
  Cambridge University Press, Cambridge, 2000.
  
  \bibitem{Born} M.~Born,  A suggestion for unifying quantum theory and relativity, Proc.~R.~Soc.~Lond. A 165  (1938) 291.
  
\bibitem{SchroersCracow} B.~J.~Schroers, Quantum gravity and non-commutative spacetimes in three dimensions: a unified approach, Acta Phys. Pol. B Proceedings Supplement vol. 4 (2011) 379--402. 



\bibitem{LNRT}J.~Lukierski, A.~Nowicki,  H.~Ruegg and V.~Tolstoi,
 $q$-deformation of Poincar\'e algebra, Phys.~Lett.~B264 (1991) 331--338.
\bibitem{LNR}J.~Lukierski, A.~Nowicki and   H.~Ruegg,
 New quantum  Poincar\'e algebra and $\kappa$-deformed field theory, Phys.~Lett.~B293 (1992) 344--352.
\bibitem{MR} S.~Majid and H.~Ruegg, Bicrossproduct structure of the 
$\kappa$-Poincar\'e group and non-commutative geometry, Phys.~Lett.~B334 (1994) 348--354.

\bibitem{relloc} G.~Amelino-Camelia, L.~Freidel, J.~Kowalski-Glikman, L.~Smolin, Relative locality: A deepening of the relativity principle, , Int.~J. Mod.~Phys.~D  20 (2011) 2867--2873, 

\bibitem{Stachura}
P.~Stachura, Poisson-Lie structures on Poincar\'e and Euclidean groups in three dimensions, J.~Phys.~A: Math. Gen. 31 (1998) 4555--4564.


\bibitem{OS1}P.~K.~Osei and B.~J.~Schroers, 
On the semiduals of local isometry groups in 3d gravity, J.~Math.~ Phys.   53  (2012) 073510. 


\bibitem{FNR} L.~Freidel, K.~Noui and  P.~Roche, 6J symbols duality relations,   J.~Math.~Phys.~48 (2007) 113512.


\bibitem{MajidSchroers}S.~Majid and B.~J.~Schroers, q-deformation and semi-dualisation in 3d quantum gravity,
J. ~Phys.~A 42 (2009) 425402.

\bibitem{Meusburger} C.~Meusburger,  Geometrical (2+1)-gravity and the
Chern-Simons formulation: Grafting, Dehn twists, Wilson loop
observables and the cosmological constant, Commun.~Math.~Phys.~273 (2007) 705--754.

 \bibitem{MS}  C.~Meusburger and B.~J.~Schroers, 
Quaternionic and Poisson-Lie structures in 3d gravity: the cosmological 
constant as deformation parameter,
J.~Math.~Phys.~49 (2008) 083510.


\bibitem{MSkappa}  C.~Meusburger and B.~J.~Schroers, Generalised Chern-Simons actions for 3d gravity and kappa-Poincare symmetry, Nucl. Phys. B 806 (2009) 462--488.


\bibitem{Taussky} O.~Taussky, The factorisation of the adjugate of a finite matrix, Linear Algebra Appl. 1 (1968) 39--41. 


\bibitem{BHDS} A.~Ballesteros, F.~J.~Herranz, M.~A.~Del Olmo and M.~Santander,
Quantum (2+1) kinematical algebras: a global approach, 
J.~Phys.~A: Math.~Gen.~ 27 (1994) 1283--1298.


\bibitem{BHDSspacelikedeformation} A.~Ballesteros, F.~J.~Herranz, M.~A.~Del Olmo and M.~Santander, 	
4-D quantum affine algebras and space-time q-symmetries,
 J.~Math.~Phys. 35 (1994) 4928--4940.


\bibitem{BHDSlightlikedeformation} A.~Ballesteros, F.~J.~Herranz, M.~A.~Del Olmo and M.~Santander, A new null-plane quantum Poincar\'e algebra, 
Phys.~Lett.~B351 (1995) 137--145.

\bibitem{Hall} G.~S.~Hall, Symmetries and Curvature in General Relativity, World Scientific Publishing, 2004
\end{thebibliography}
\end{document}